\documentclass{aastex631}

\begin{document}

\title{X-ray observations of the Zwicky 3146 galaxy cluster reveal a 3.5~keV excess}

\author[0000-0003-3851-7219]{Sunayana Bhargava}
\affiliation{Universit\'e C\^ote d'Azur, Observatoire de la C\^ote d'Azur, CNRS, Laboratoire Lagrange, Nice, France} 

\author{Paul Giles}
\affiliation{Department of Physics and Astronomy, University of Sussex, Falmer, Brighton BN1 9QH, UK}

\author{Kathy Romer}
\affiliation{Department of Physics and Astronomy, University of Sussex, Falmer, Brighton BN1 9QH, UK}

\author{Tesla Jeltema}
\affiliation{Santa Cruz Institute for Particle Physics, University of California, Santa Cruz, 1156 High St, Santa Cruz, CA 95064, USA}

\author{Devon Hollowood}
\affiliation{Santa Cruz Institute for Particle Physics, University of California, Santa Cruz, 1156 High St, Santa Cruz, CA 95064, USA}

\author{Matt Hilton}
\affiliation{Wits Centre for Astrophysics, School of Physics, University of the Witwatersrand, Private Bag 3, 2050, Johannesburg, South Africa}
\affiliation{Astrophysics Research Centre, School of Mathematics, Statistics, and Computer Science, University of KwaZulu-Natal, Westville Campus, Durban 4041, South Africa}



\begin{abstract}

In this note, we present spectral fits of the well-documented sloshing cool-core cluster Zwicky 3146 ($z=0.291$), to test the existence of the highly speculated 3.5~keV line. We report excesses at $>3\sigma$ significance at $E=3.575$~keV, yielding a flux $F = 8.73_{-2.22}^{+2.17}$ $\times 10^{-6}$ photons cm$^{-2}$ s$^{-1}$, in \textit{XMM-Newton}, and $E=3.55$~keV, with a flux $F = 10.0_{-2.96}^{+3.05}$ $\times 10^{-6}$ photons cm$^{-2}$ s$^{-1}$ in \textit{Chandra}. We explore the possibility that the 3.5 keV excess is correlated to the presence of cold gas within the cluster, based on optical and sub-mm literature analyses. Following the launch of the X-ray Imaging and Spectroscopy Mission (XRISM), high resolution spectroscopy ($\leq 7$ eV) will reveal in unprecedented detail, the origin of this unidentified feature, for which Zwicky 3146 should be considered a viable target, due to the strength of the feature in two independent X-ray telescopes, opening a new window into plasma or charge exchange studies in galaxy clusters. 

\end{abstract}

\section{Introduction} \label{sec:intro}

In 2014, reports of an unidentified flux excess in the stacked spectra of galaxy clusters invigorated the X-ray astronomy community as a possible signature of a sterile neutrino dark matter \citep[][hereafter B14]{Bulbul2014}, whose decay processes are hypothetically detectable by X-ray telescopes \citep{Abazajian2001}. Subsequent studies provided confirmed and tentative detections e.g. \cite{JeltemaProfumo2015} and null results e.g. \cite{Sicilian2020,Dessert2024}. In work by \cite{Bhargava2020}, hereafter B20, a global excess from 114 clusters was ruled out at the $3\sigma$ level, however, three clusters individually exhibited strong excesses in the 3.5 keV range which were not explainable by either dark matter or modelling interpretations. An alternative interpretation of the feature is charge exchange \citep[hereafter CX, see][]{Gu2015ChargeExchange}{}{}. This happens when a neutral atom collides with a sufficiently charged ion, and then recombines the ion into a highly excited state. In the context of galaxy clusters, this process occurs when cold gas (e.g. hydrogen) meets a highly charged ion in the ICM. Such cold clouds can exist in central galaxies, or infalling cluster members, whose cold gas has not yet been removed by ram pressure stripping, making cool core clusters an ideal starting point for such a phenomenon. While the CX mechanism has been observed in numerous astrophysical contexts, CX has so far never been clearly detected in galaxy clusters. One galaxy cluster that is a viable target for the CX mechanism is Zwicky 3146. This cluster is chosen for having among the most well-studied properties, notably the discovery of a strong cooling flow \citep{Egami2006} and strikingly large quantities of molecular hydrogen ($5 \times 10^{10} M_{\odot}$) in the brightest cluster galaxy measured using ALMA observations \citep{Vantyghem2021}. Moreover, this system has the largest individual XMM exposure time in the stacked sample of clusters from the seminal work of B14, in which a line was detected at high significance, and multiple \textit{Chandra} observations.

\section{Methods} \label{sec:method}

For both available XMM and \textit{Chandra} observations, we extract the spectrum of the cluster and fit as follows. Prior to fitting the X-ray spectrum, each one is blueshifted (i.e. so that $z_{\rm{effective}}=0$). While this step is not strictly necessary for individual clusters, we do it to maintain consistency with the methodology in B20 applied to individual clusters, and to retain the rest-frame definition of the 3.5 keV line. The format of a source spectrum measured by the detector is a list of photon counts as a function of channel number. The associated cluster response matrix file ({\tt RMF}) and ancillary response file ({\tt ARF}) contain the energy ranges corresponding to the source spectrum channels. Each cluster spectrum is blueshifted by rescaling the upper and lower energy bounds for each photon channel by a factor of $1+z$. Next, we carry out a fit to a fiducial model (`model A': {\tt tbabs $\times$ apec}) and then compare the goodness of fit to a model that includes an addition emission line component (`model B': {\tt tbabs $\times$ (apec + Gaussian)}) to mimic an unknown line feature. The fitting is performed using {\sc xspec} version {\sc 12.10.1f}, {\sc apec} version 3.0.9, using the {\sc xspec} {\tt cstat} statistic. There are five parameters in model A. Three are frozen during the fit: the $n_{\rm H}$ value, the X-ray temperature (at the $T_{\rm X}^{\rm PN}$ value), and the redshift (at $z_{\rm effective}=0$). Details of the X-ray temperature measurement are described in B20 for XMM and \cite{Hollowood2019}, for \textit{Chandra}. Two are left free: the {\tt apec} normalisation, and the metal abundance. There are nine parameters in model B. Five of these are shared with model A and treated in the same way during the fit. The remaining three parameters are associated with the Gaussian component: the central energy, line width, normalisation. The central energy is frozen at a value iterated between $3 - 5$~keV in intervals of 25 eV. The line width is fixed at zero to mimic the narrowest possible line emission allowed by the energy resolution of the detector, which is in turn defined by the {\tt ARF} matrix associated with the cluster spectrum. The normalisation is a free parameter. We define the parameter $\Delta C$ to quantify the goodness of fit between the two models at a given energy $E$, where $3<E<5$~keV (see above). $\Delta C$ is the difference between the value of the Cash statistic after fitting for model A and the value after fitting for model B. A positive value of $\Delta C$ indicates that the fit is better for model B. The estimate for the $3\sigma$ threshold (i.e. the value of $\Delta C$ above which is considered a significantly better fit), is calculated based on the probability of exceeding 99$\%$ significance for model B compared to model A, taking into account the fact that model B has one additional degree of freedom. 

\section{Results and Conclusions} \label{sec:floats}

\begin{figure}[ht!]
\plotone{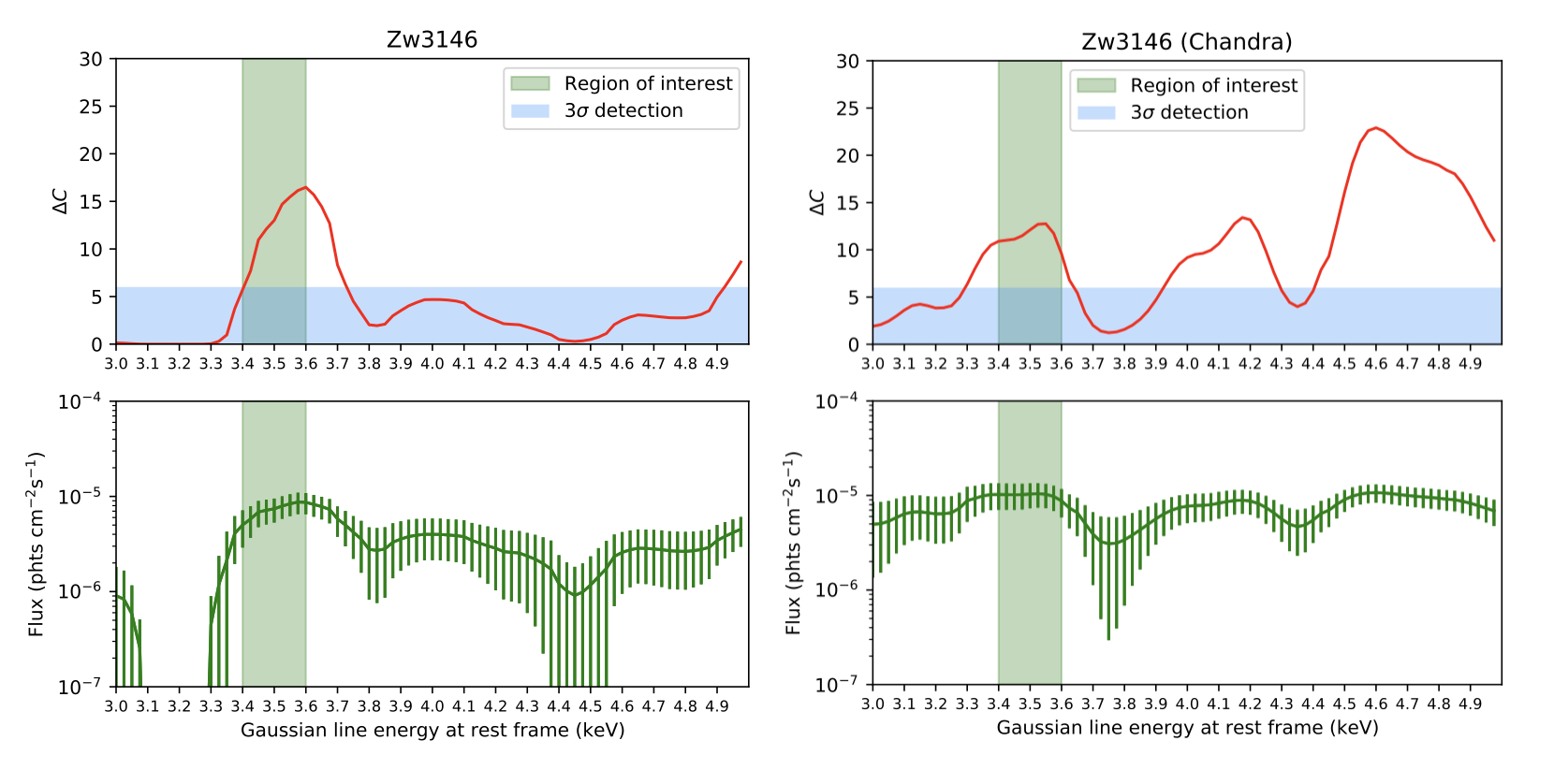}
\caption{Single temperature model fits from \textit{XMM-Newton} (left) and \textit{\textit{Chandra}} (right) spectra for Zw3146 that show a fit improvement above $3\sigma$ (blue region) in the energy range of interest (green shaded region). For each plot, the top panel (red line) shows the change in fit statistic $\Delta C$ as a function of energy in the range $3-5$~keV. The bottom panels (green line) show the fitted normalisation of the Gaussian line and corresponding errors. The XMM observation ID is 0605540301 (cleaned exp time 34.1 ks); the \textit{Chandra} observation is 9371 (cleaned exp time 39.7 ks).   
\label{fig:zw3146}}
\end{figure}

We fit the XMM and \textit{Chandra} spectra of Zwicky 3146 using the methods described in Section \ref{sec:method}. The results, shown in Fig. \ref{fig:zw3146} display the 3.5~keV excess in both spectra in the form of a positive $\Delta C$ relative to the fiducial model, using the same assessment of statistical significance in B20. The presence of a feature in both instruments implies that the excess is not due to an instrumental artefact. We also note that in the case of the \textit{Chandra} spectral fit (Fig. \ref{fig:zw3146}, right), there are two further excesses at 4.4~keV and 4.7~keV, which arise from a blend of known instrumental lines and the astrophysical Ca XIX transition lines, respectively \citep{Cappelluti2018}. We also note that for a cluster such as Zwicky 3146, which contains multiphase gas, we make a simplifying assumption by assuming a single temperature for the system. Multi-temperature fits extracted in different regions of Zwicky 3146 have been performed in e.g. \cite{Allen1996}. Finally, we do not include fitting of any core-excluded spectra for either the \textit{XMM-Newton} or \textit{Chandra} spectra, as previous studies such as B20 did not find the presence of an excess to be correlated to core exclusion. 

The sterile neutrino dark matter mixing angle (computed from Section 3.3 of B20), derived from the 3.5~keV flux estimate is incompatible with previous and current constraints summarised in \cite{Fong2024}. The diminishing possibility of a dark matter origin for the line suggests that novel gas mechanisms such as CX could be responsible. Such an interpretation would be consistent with the appearance of a strong feature in individual clusters, while not being a global property or increasing with cluster mass. As XRISM prepares to observe some of its first galaxy cluster targets in the months ahead, such high resolution spectroscopy in the soft X-ray band will be critical in understanding the role of multiphase gas interactions in the ICM. The cluster Zwicky 3146, showing a detection from both XMM and \textit{Chandra} at the current resolution level, is therefore a valuable target for XRISM, in addition to other clusters such Perseus, Virgo, Coma, and Centaurus. 

\section{Acknowledgments}
SB would like to thank L. Gu and E. Koulouridis for valuable discussions and guidance. This paper uses a Chandra observation, obtained by the Chandra X-ray Observatory, contained in~\dataset[doi:10.25574/cdc.237]{https://doi.org/10.25574/cdc.237}.

%


\bibliography{sample631}{}

\begin{thebibliography}{}
\expandafter\ifx\csname natexlab\endcsname\relax\def\natexlab#1{#1}\fi
\providecommand{\url}[1]{\href{#1}{#1}}
\providecommand{\dodoi}[1]{doi:~\href{http://doi.org/#1}{\nolinkurl{#1}}}
\providecommand{\doeprint}[1]{\href{http://ascl.net/#1}{\nolinkurl{http://ascl.net/#1}}}
\providecommand{\doarXiv}[1]{\href{https://arxiv.org/abs/#1}{\nolinkurl{https://arxiv.org/abs/#1}}}

\bibitem[{{Abazajian} {et~al.}(2001){Abazajian}, {Fuller}, \& {Patel}}]{Abazajian2001}
{Abazajian}, K., {Fuller}, G.~M., \& {Patel}, M. 2001, \prd, 64, 023501, \dodoi{10.1103/PhysRevD.64.023501}

\bibitem[{Allen {et~al.}(1996)Allen, Fabian, Edge, Bautz, Furuzawa, \& Tawara}]{Allen1996}
Allen, S., Fabian, A., Edge, C., {et~al.} 1996, MNRAS, 283, \dodoi{10.1093/mnras/283.1.263}

\bibitem[{{Bhargava} {et~al.}(2020){Bhargava}, {Giles}, {Romer}, {Jeltema}, {Mayers}, {Bermeo}, {Hilton}, {Wilkinson}, {Vergara}, {Collins}, {Manolopoulou}, {Rooney}, {Rosborough}, {Sabirli}, {Stott}, {Swann}, \& {Viana}}]{Bhargava2020}
{Bhargava}, S., {Giles}, P.~A., {Romer}, A.~K., {et~al.} 2020, \mnras, 497, 656, \dodoi{10.1093/mnras/staa1829}

\bibitem[{{Bulbul} {et~al.}(2014){Bulbul}, {Markevitch}, {Foster}, {Smith}, {Loewenstein}, \& {Randall}}]{Bulbul2014}
{Bulbul}, E., {Markevitch}, M., {Foster}, A., {et~al.} 2014, \apj, 789, 13, \dodoi{10.1088/0004-637X/789/1/13}

\bibitem[{{Cappelluti} {et~al.}(2018){Cappelluti}, {Bulbul}, {Foster}, {Natarajan}, {Urry}, {Bautz}, {Civano}, {Miller}, \& {Smith}}]{Cappelluti2018}
{Cappelluti}, N., {Bulbul}, E., {Foster}, A., {et~al.} 2018, \apj, 854, 179, \dodoi{10.3847/1538-4357/aaaa68}

\bibitem[{{Dessert} {et~al.}(2024){Dessert}, {Foster}, {Park}, \& {Safdi}}]{Dessert2024}
{Dessert}, C., {Foster}, J.~W., {Park}, Y., \& {Safdi}, B.~R. 2024, \apj, 964, 185, \dodoi{10.3847/1538-4357/ad2612}

\bibitem[{{Egami} {et~al.}(2006){Egami}, {Rieke}, {Fadda}, \& {Hines}}]{Egami2006}
{Egami}, E., {Rieke}, G.~H., {Fadda}, D., \& {Hines}, D.~C. 2006, \apjl, 652, L21, \dodoi{10.1086/509886}

\bibitem[{{Fong} {et~al.}(2024){Fong}, {Ng}, \& {Liu}}]{Fong2024}
{Fong}, C., {Ng}, K. C.~Y., \& {Liu}, Q. 2024, arXiv e-prints, arXiv:2401.16747, \dodoi{10.48550/arXiv.2401.16747}

\bibitem[{{Gu} {et~al.}(2015){Gu}, {Kaastra}, {Raassen}, {Mullen}, {Cumbee}, {Lyons}, \& {Stancil}}]{Gu2015ChargeExchange}
{Gu}, L., {Kaastra}, J., {Raassen}, A.~J.~J., {et~al.} 2015, \aap, 584, L11, \dodoi{10.1051/0004-6361/201527634}

\bibitem[{{Hollowood} {et~al.}(2019){Hollowood}, {Jeltema}, {Chen}, {Farahi}, {Evrard}, {Everett}, {Rozo}, {Rykoff}, {Bernstein}, {Bermeo-Hernandez}, {Eiger}, {Giles}, {Israel}, {Michel}, {Noorali}, {Romer}, {Rooney}, \& {Splettstoesser}}]{Hollowood2019}
{Hollowood}, D.~L., {Jeltema}, T., {Chen}, X., {et~al.} 2019, \apjs, 244, 22, \dodoi{10.3847/1538-4365/ab3d27}

\bibitem[{{Jeltema} \& {Profumo}(2015)}]{JeltemaProfumo2015}
{Jeltema}, T., \& {Profumo}, S. 2015, \mnras, 450, 2143, \dodoi{10.1093/mnras/stv768}

\bibitem[{{Sicilian} {et~al.}(2020){Sicilian}, {Cappelluti}, {Bulbul}, {Civano}, {Moscetti}, \& {Reynolds}}]{Sicilian2020}
{Sicilian}, D., {Cappelluti}, N., {Bulbul}, E., {et~al.} 2020, \apj, 905, 146, \dodoi{10.3847/1538-4357/abbee9}

\bibitem[{{Vantyghem} {et~al.}(2021){Vantyghem}, {McNamara}, {O'Dea}, {Baum}, {Combes}, {Edge}, {Fabian}, {McDonald}, {Nulsen}, {Russell}, \& {Salom{\'e}}}]{Vantyghem2021}
{Vantyghem}, A.~N., {McNamara}, B.~R., {O'Dea}, C.~P., {et~al.} 2021, \apj, 910, 53, \dodoi{10.3847/1538-4357/abe306}

\end{thebibliography}
\bibliographystyle{aasjournal}



\end{document}